\documentclass[prl,floatfix,superscriptaddress,preprint]{revtex4}
\usepackage{latexsym}
\usepackage{amstext,amsfonts}
\usepackage{epsfig}

\begin{document}

\title{Limited path percolation in complex networks}

\author{Eduardo L\'{o}pez}
\thanks{These authors contributed equally to this work}
\affiliation{Center for Non-Linear Studies \& T-13, Theoretical Division, Los Alamos National
Laboratory, Los Alamos, NM 87545, USA}
\author{Roni Parshani}
\thanks{These authors contributed equally to this work}
\affiliation{Minerva Center \& Department of Physics, Bar-Ilan University, Ramat Gan, Israel}
\author{Reuven Cohen}
\affiliation{New England Complex Systems Institute, Cambridge, MA, USA}
\affiliation{Massachusetts Institute of Technology, Cambridge, MA, USA}
\author{Shai Carmi}
\affiliation{Minerva Center \& Department of Physics, Bar-Ilan University, Ramat Gan, Israel}
\author{Shlomo Havlin}
\affiliation{Minerva Center \& Department of Physics, Bar-Ilan University, Ramat Gan, Israel}
\date{February 14, 2007}

\begin{abstract}
We study the stability of network communication after removal of $q=1-p$ links under the 
assumption that communication is effective only if the shortest path between nodes 
$i$ and $j$ after removal is shorter than $a\ell_{ij} (a\geq1)$ where $\ell_{ij}$ 
is the shortest path before removal. 
For a large class of networks, we find a new percolation transition 
at $\tilde{p}_c=(\kappa_o-1)^{(1-a)/a}$, where 
$\kappa_o\equiv \langle k^2\rangle/\langle k\rangle$ and 
$k$ is the node degree.
Below $\tilde{p}_c$, only a fraction $N^{\delta}$ of the network nodes can communicate, 
where $\delta\equiv a(1-|\log p|/\log{(\kappa_o-1)}) < 1$,
while above $\tilde{p}_c$, order $N$ nodes can communicate
within the limited path length $a\ell_{ij}$. Our analytical results 
are supported by simulations on Erd\H{o}s-R\'{e}nyi and scale-free network models.
We expect our results to influence the design of networks, routing algorithms, 
and immunization strategies, where short paths are most relevant.
\end{abstract}
\maketitle

The study of complex networks has emerged as an important tool to better
understand many social, technological, and biological real-world systems
ranging from communication networks like the Internet, to cellular
networks~\cite{rev-Albert}.
In many cases, networks are the medium through which information is transported,
i.e., in social networks the propagation of epidemics, rumors, etc.
and in the Internet the propagation of data 
packets~\cite{Wasserman,Pastor-Satorras,Newman,Lopez,Sreenivasan}.

An important question regarding networks is their stability, i.e, under 
what conditions the network breaks
down~\cite{Barabasi,Cohen,Callaway,Tanizawa}. 
In communications, a network breakdown means information cannot be 
transmitted to most nodes, and in epidemiology, that an epidemic has stopped.

The main approach for studying network stability is percolation theory~\cite{Percolation}.
In percolation, a fraction $q=1-p$ of the $N$ network nodes (or links) 
are removed until a critical value $p_c$ is reached. For $p< p_c$ 
the network collapses into small clusters,
while for $p>p_c$, a spanning cluster of order $N$ nodes 
appears~\cite{Cohen,Callaway,Percolation,ER,Bollobas}. 
However, even though in the original network the nodes are connected
through short paths, near $p_c$ the paths become very long.
For instance, in the original Erd\H{o}s-R\'{e}nyi network
the typical distance between nodes is of order $\log N$~\cite{ER}
compared to order $N^{1/3}$ near the percolation threshold~\cite{Braunstein}.
These long distances may have a significant influence on network function.
For example, in communication, long paths are usually inefficient, and in 
epidemics, disease spreading 
often decays in time due to mutations or natural immunization, so for 
long paths the epidemic may die out before the network collapses. 
In these cases the interesting question is sometimes, not when does the 
network break down, but when the network connectivity becomes inefficient.

To answer this question, we propose a new percolation model which we call 
limited path percolation (LPP). In this model, after removing a fraction $q=1-p$ 
of the network nodes, any two of these nodes, say $i$ and $j$ are considered connected 
only if the shortest path between them is shorter than $a\ell_{ij}$ ($a\geq 1$), where 
$\ell_{ij}$ is the shortest path before removal. We then ask, given our new limited 
path constrains,
what is the value $p$ at which a spanning cluster appears. We find a new 
phase transition, which depends on $a$, at $\tilde{p}_c\equiv \tilde{p}_c(a)$,
where $p_c<\tilde{p}_c<1$. For $p_c <p< \tilde{p}_c$, the LPP 
spanning cluster is only a zero fraction (fractal) of the network, which scales as $N^{\delta}$
($\delta<1$). For $p> \tilde{p}_c$ the LPP spanning cluster is of order $N$. 

For simplicity, we start our analysis with Erd\H{o}s-R\'{e}nyi (ER) 
networks and then argue that the theory is also valid in general for 
random networks.
We begin with random removal but
extend our considerations to targeted removal on highly connected nodes,
and find that similar phenomena appears. We support our theory with simulations.

Erd\H{o}s-R\'{e}nyi networks \cite{ER,Bollobas} are random networks consisting of $N$
nodes connected with probability $\phi$ and disconnected with probability $1-\phi$.
The degree distribution $\Phi(k)$ is Poisson with the form
$\Phi(k)=\langle k\rangle^k e^{-\langle k\rangle}/k!$,
where $k$, the degree, is the number of links attached to a node, 
and $\langle k\rangle\equiv\sum_{k=1}^{\infty} k\Phi(k)$ is the average
degree of the network. The typical distance between nodes is 
$\log N/\log\langle k\rangle$.

\begin{figure}
\begin{center}
\epsfig{file=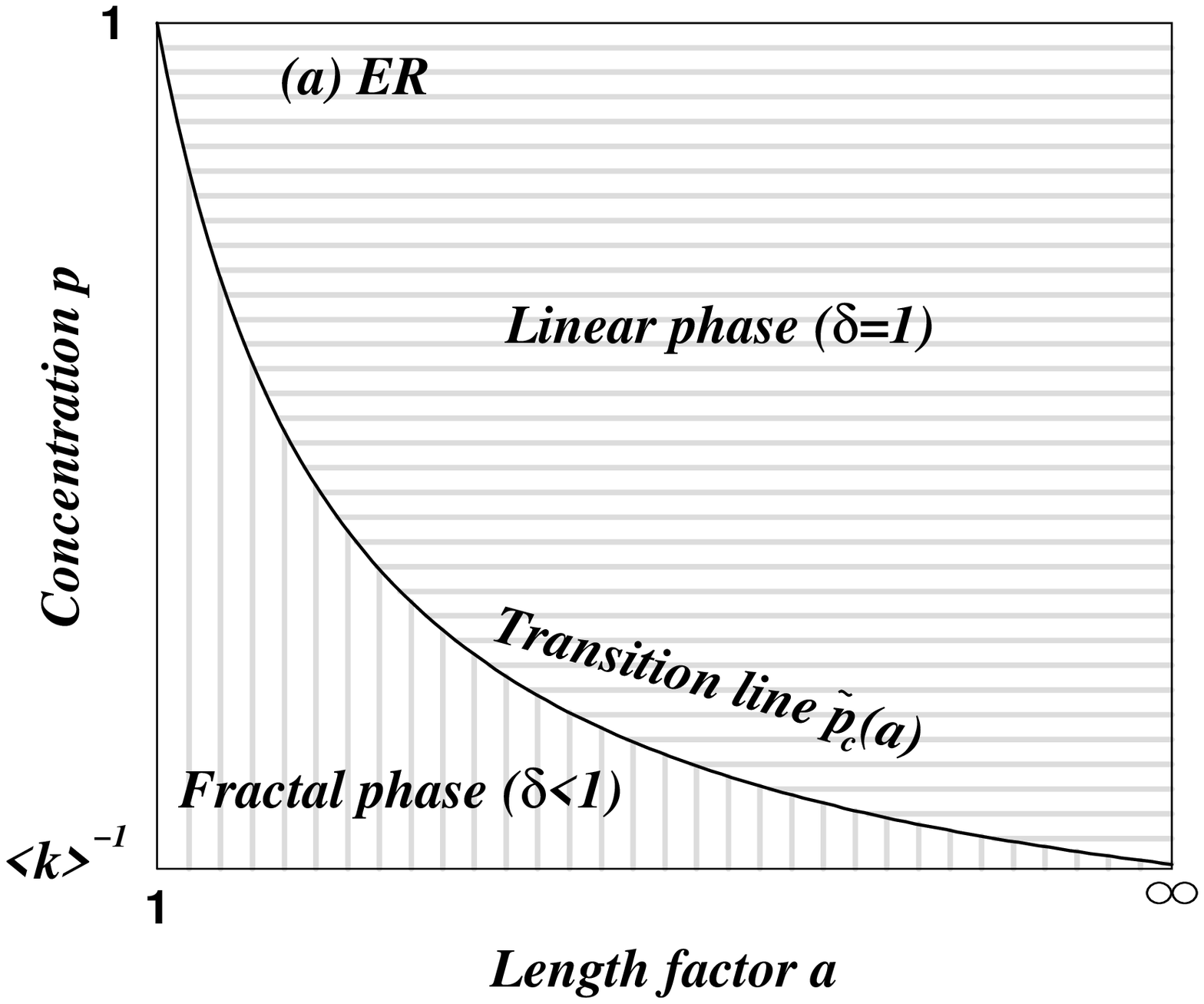,height=7cm,width=8cm}
\epsfig{file=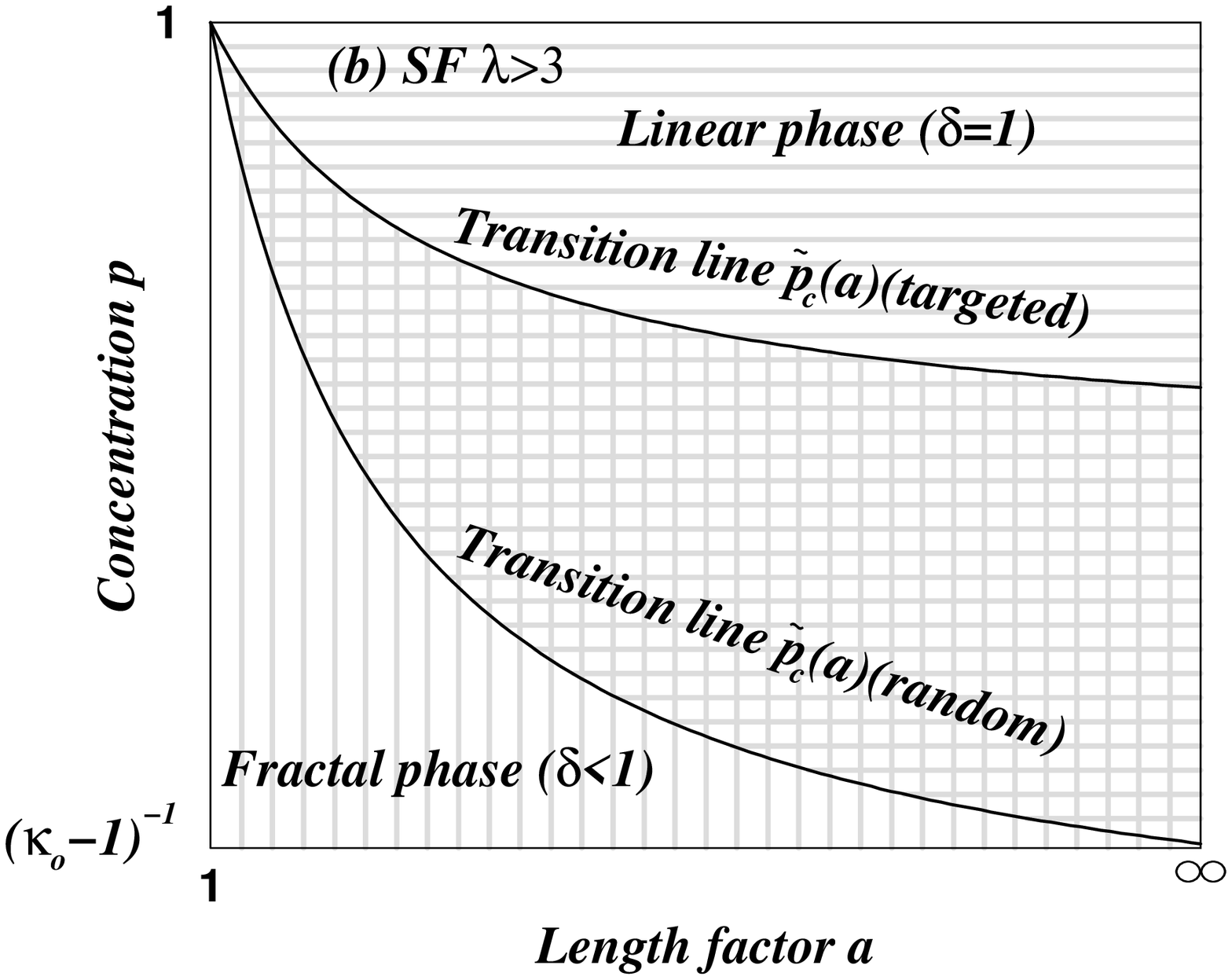,height=7cm,width=8cm}
\end{center}
\caption{(a) Phase diagram for Erd\H{o}s-R\'{e}nyi networks 
of LPP with respect to parameters $a$ and $p$, demonstrating the linear and 
power law (fractal) phases for $S_{a}\sim N^{\delta}$. 
(b) Similar phase diagram for scale-free networks with $\lambda>3$. 
The two transition lines represent networks with the same $\kappa$. Note the 
slow decrease
of the transition line for targeted removal compared to the transition line for
random removal. The region between the two lines has a power law (fractal) phase for 
targeted removal and a linear phase for random removal.
In both (a) and (b) the regular percolation threshold is fiven by the limit
$a\rightarrow\infty$, i.e, $p_c=\langle k\rangle^{-1}$ for ER and
$p_c=(\kappa_o-1)^{-1}$ for SF with $\lambda>3$.
}
\label{phase_diagram}
\end{figure}

\begin{figure}[t]
\begin{center}
\epsfig{file=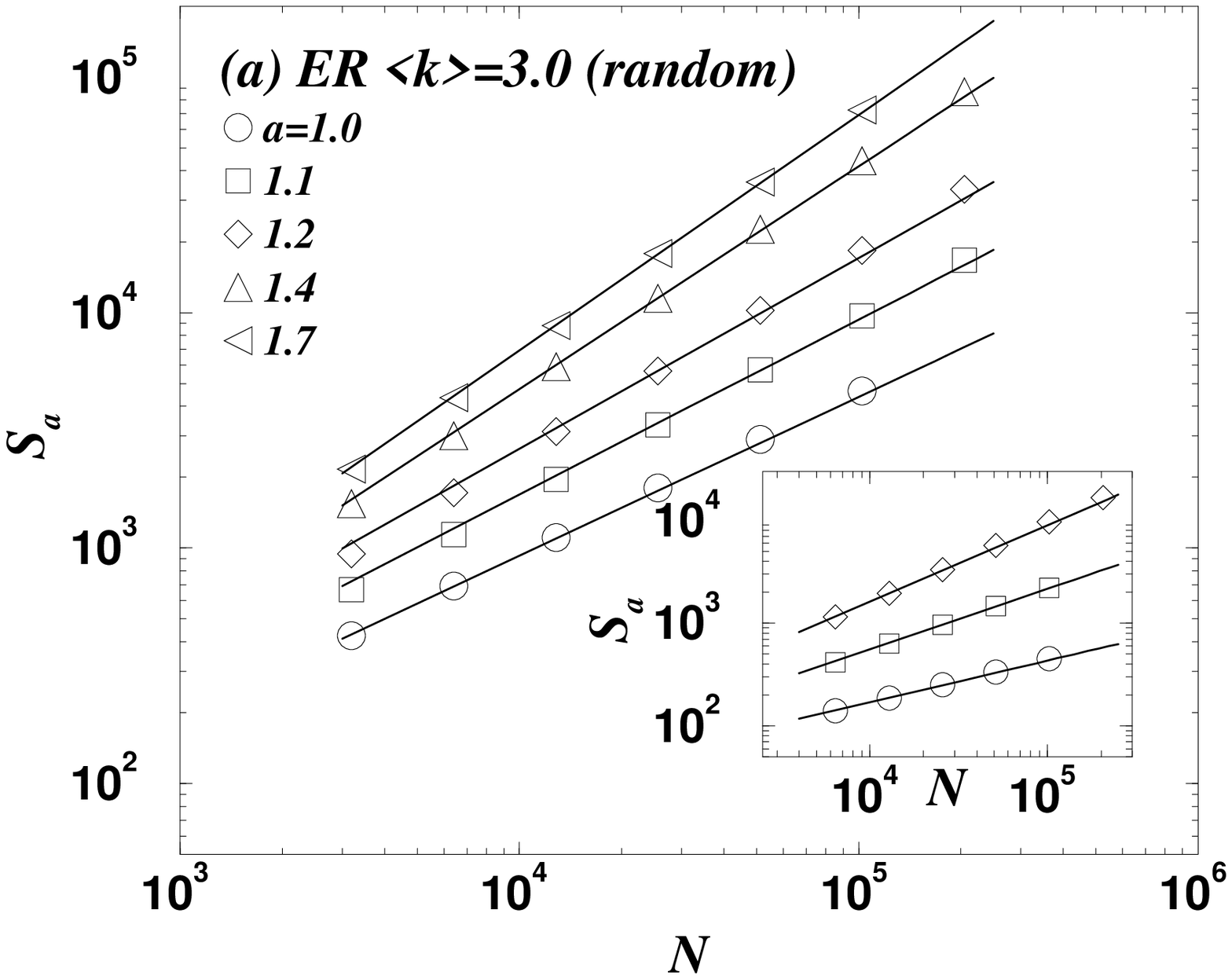,height=7cm,width=8cm}
\epsfig{file=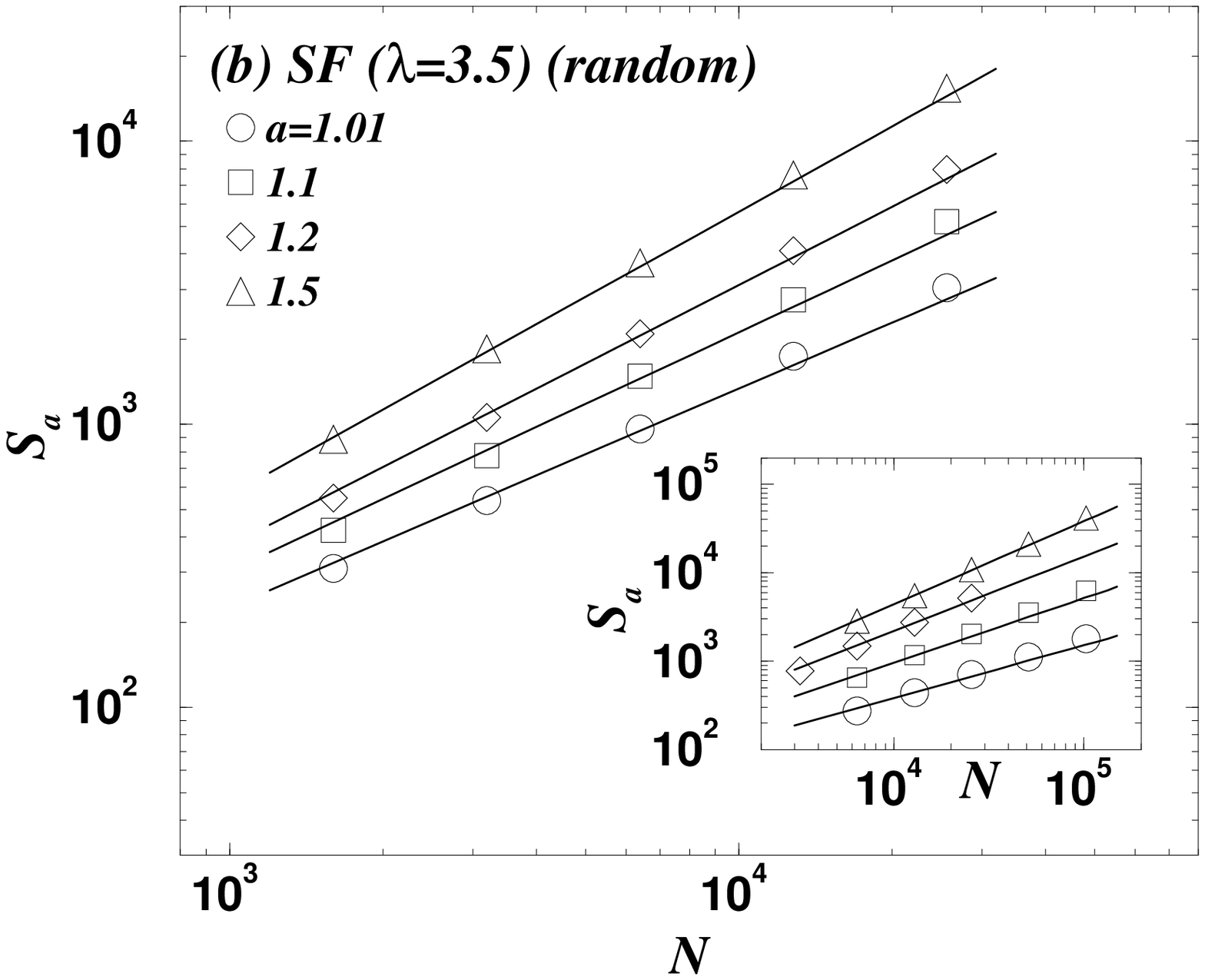,height=7cm,width=8cm}
\epsfig{file=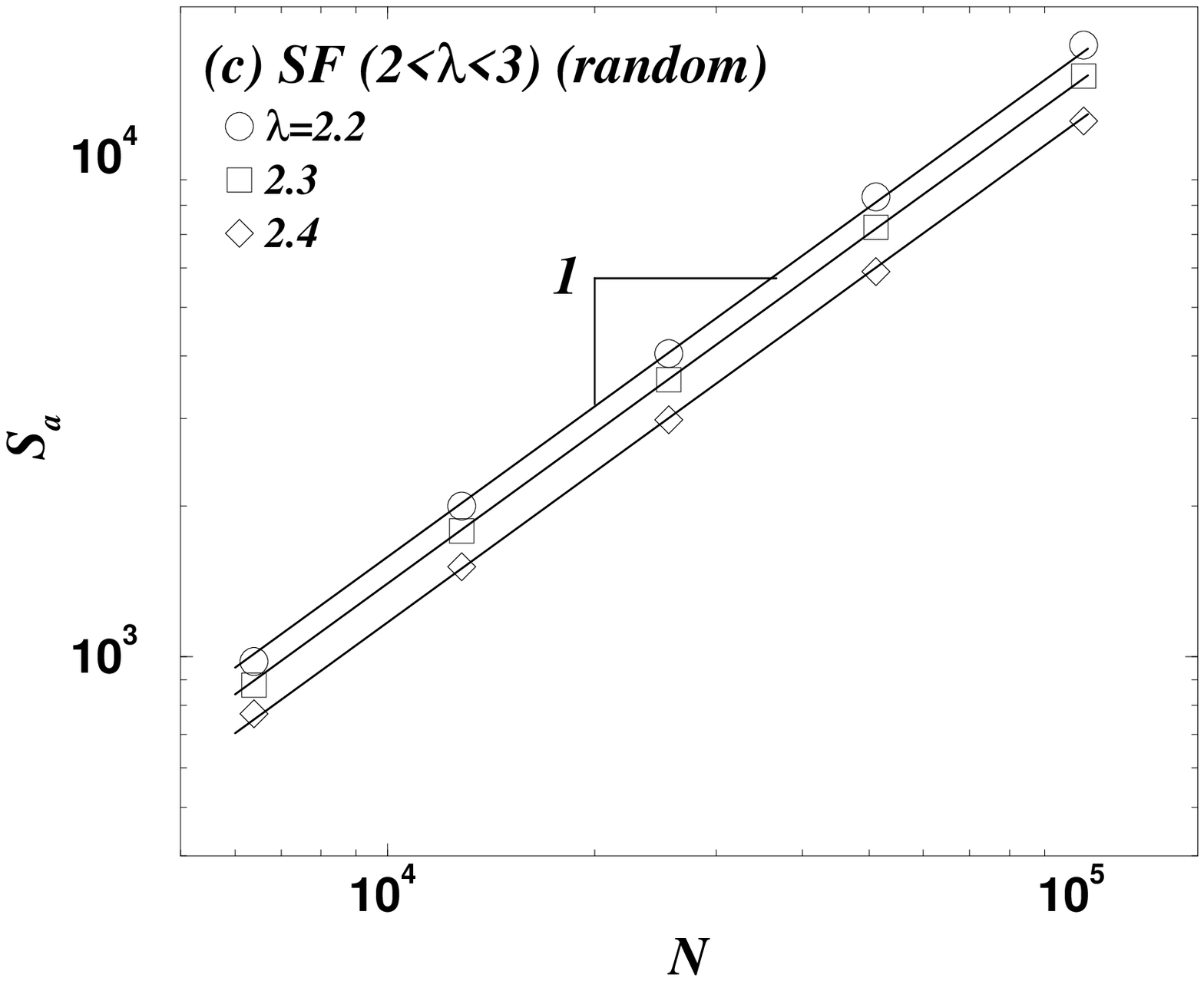,height=7cm,width=8cm}
\epsfig{file=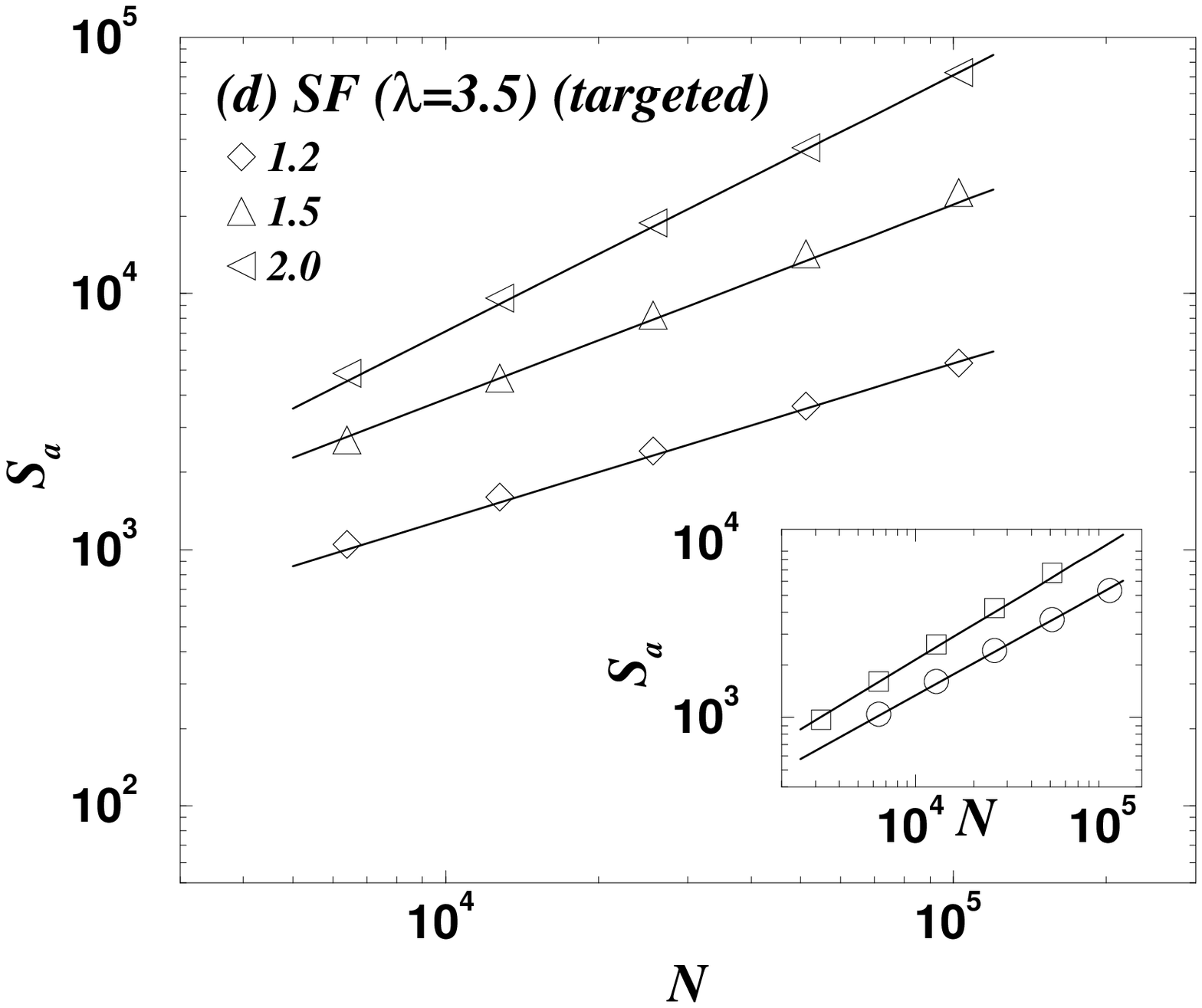,height=7cm,width=8cm}
\end{center}
\caption{Simulation results (symbols) for $S_a$ vs. $N$ for various network types
under random or targeted removal and different values of $a$ and $p$
(indicated in plot legends),
compared to the theoretically predicted power laws (solid lines), with $\delta$ 
calculated from Table~\ref{table1}. Network sizes are typically between 1600 and 204800.
In all plots the simulation results agree with the theoretical predictions.
(a) ER networks (random) with $\langle k\rangle= 3$ for fixed $p=0.7$ and different $a$ 
values.
Inset shows the same networks with fixed $a=1.1$ and $p=0.5 (\bigcirc), 0.6 (\Box)$
and 0.7 ($\diamondsuit$).
(b) SF networks (random) with $\lambda=3.5$, $m=2$, $p=0.7$ and different $a$ values.
Inset shows the same networks with fixed $a=1.1$ and $p=0.5 (\bigcirc), 0.6 (\Box),
0.7 (\diamondsuit)$ and 0.8 ($\bigtriangleup$).
(c) SF networks (random), $\lambda=2.2,2.3$ and 2.4, $m=3$, $p=0.4$, and $a=1$.
(d) SF networks (targeted) with $\lambda=3.5$, $m=3$, fixed $p=0.92$ and different $a$
values.
Inset shows the same networks with fixed $a=1.2$ and $p=0.92 (\bigcirc)$ and
0.94 ($\Box$).
}
\label{S_a_p_N}
\end{figure}

Next we evaluate $S_a$, the size of the spanning cluster under LPP.
After the removal of fraction $q$ of the links, the spanning cluster can be considered 
tree-like since, up to order $N$, loops are negligible~\cite{Cohen}.
Thus, $S_a$ can be approximated by
\begin{equation}
S_a\sim c(p)[p\langle k\rangle]^{a\frac{\log{N}}{\log{\langle k\rangle}}}=
c(p)N^{\delta}, \qquad
\delta\equiv a\left(1-\frac{|\log{p}|}{\log{\langle k\rangle}}\right) \leq  1
\qquad \text{(Erd\H{o}s-R\'{e}nyi)}
\label{S_N_ER_delta}
\end{equation}
where $p\langle k\rangle$ is the average degree after removal, 
$c(p)\equiv c_o p\langle k\rangle/(p\langle k\rangle-1)$~\cite{fn_cp},
and $a\log{N}/\log{\langle k\rangle}$ is the new tree depth imposed by the 
limited path length restriction. The exponent $\delta=\delta(a,p,\langle k\rangle)$ 
is an increasing function of $a$,
i.e., for larger values of $a$ longer paths are valid and therefore more 
nodes are included in the spanning cluster, leading to
a higher value of $\delta$.
The exponent $\delta$ is bounded 
below by zero and above by 1, since $N$ is the maximum number of 
nodes available. Setting $\delta=1$ and solving for $p$ in Eq.~(\ref{S_N_ER_delta}) 
we obtain the transition threshold
\begin{equation}
\tilde{p}_c(a)=\langle k\rangle^{\frac{1-a}{a}}
\qquad \text{(Erd\H{o}s-R\'{e}nyi)}.
\label{tildep_ER_a}
\end{equation}
Figure~\ref{phase_diagram} presents the phase diagram for LPP. For 
$p_c\leq p\leq \tilde{p}_c(a)$ the spanning cluster is a fractal 
of size $N^{\delta}$ and $\delta$ continuously increases with $p$.  
For $p>\tilde{p}_c(a)$, a spanning cluster of order $N$ exists with path 
lengths $\ell'_{ij}\leq a\ell_{ij}$. Using the function $1-\tilde{p}_c(a)$ 
we are able to calculate for a given value of 
$a$, the percentage of links 
that can be removed before the network is no longer 
connected with effective paths, i.e., shorter than $a\ell_{ij}$. Note that for 
$a \rightarrow \infty$, when no path length restriction is imposed, we recover the 
usual percolation threshold $\tilde{p}_c(a\rightarrow\infty)=p_c 
=1/\langle k\rangle$~\cite{ER}. 
Equations~(\ref{S_N_ER_delta}) and (\ref{tildep_ER_a}) are supported by 
the simulations presented in Fig.~\ref{S_a_p_N}(a)~\cite{fn_network_algorithm,fn_sim}.
For a summary of the various equations in the article, see Table~\ref{table1}.

\begin{figure}
\begin{center}
\epsfig{file=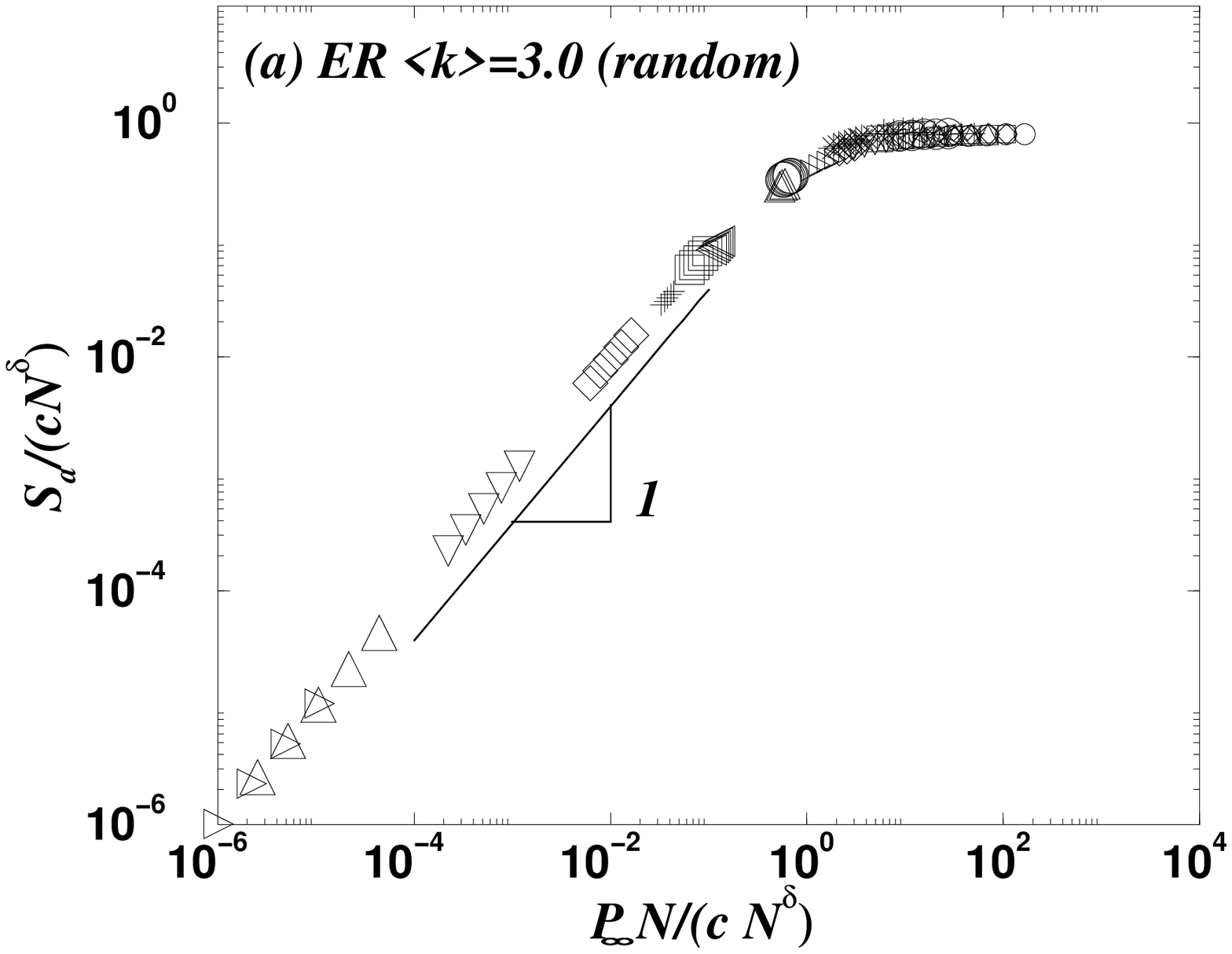,height=7cm,width=8cm}
\epsfig{file=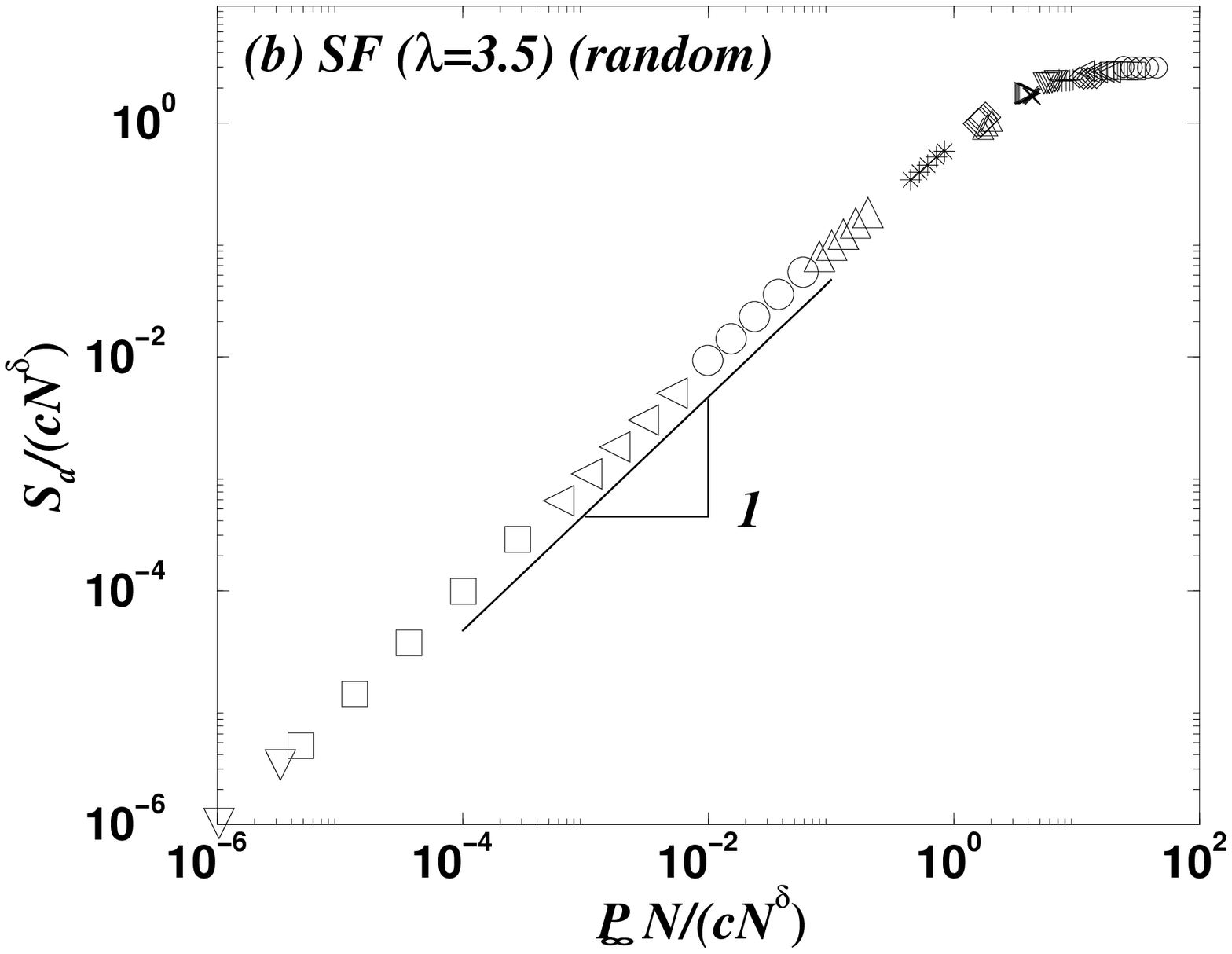,height=7cm,width=8cm}
\epsfig{file=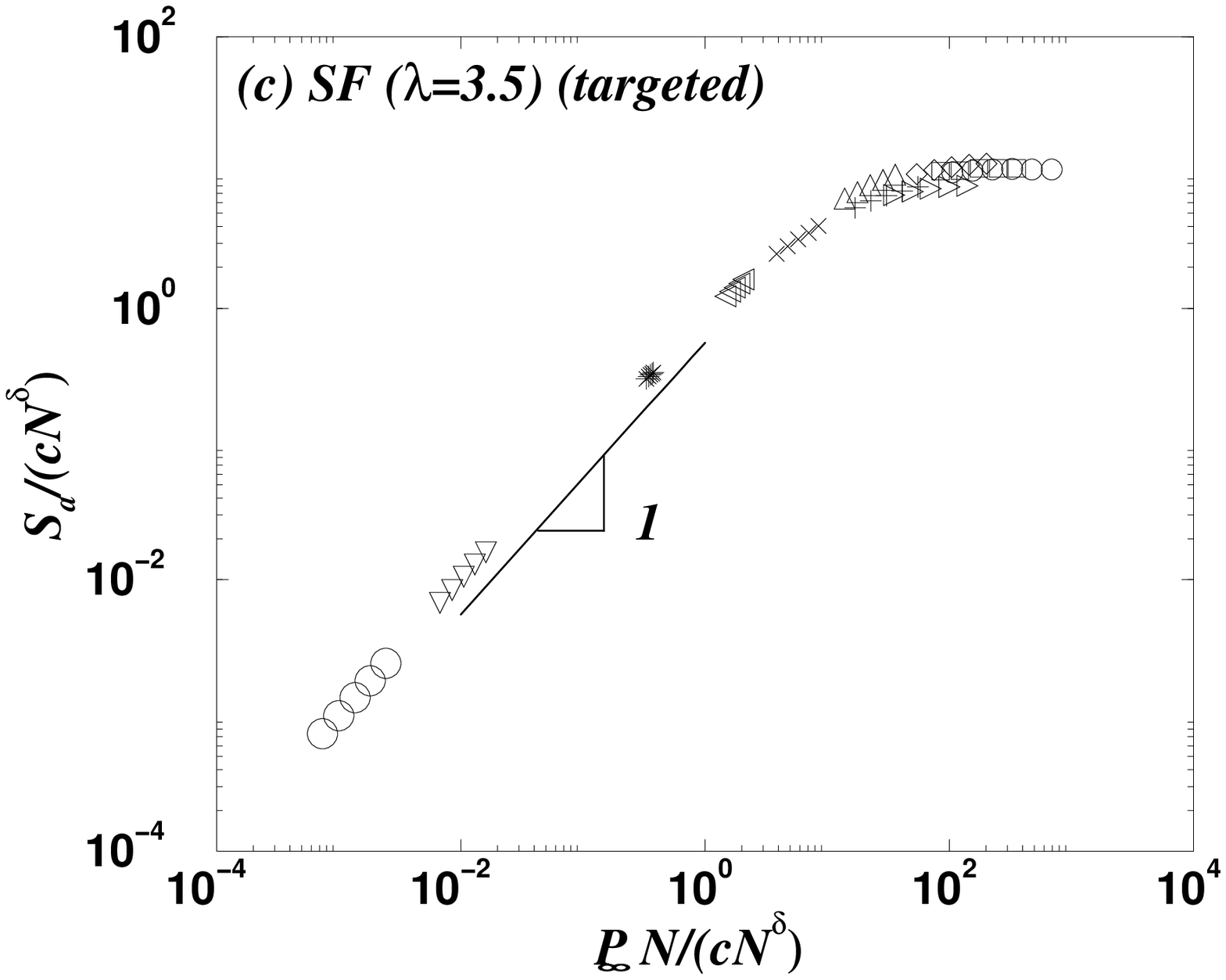,height=7cm,width=8cm}
\epsfig{file=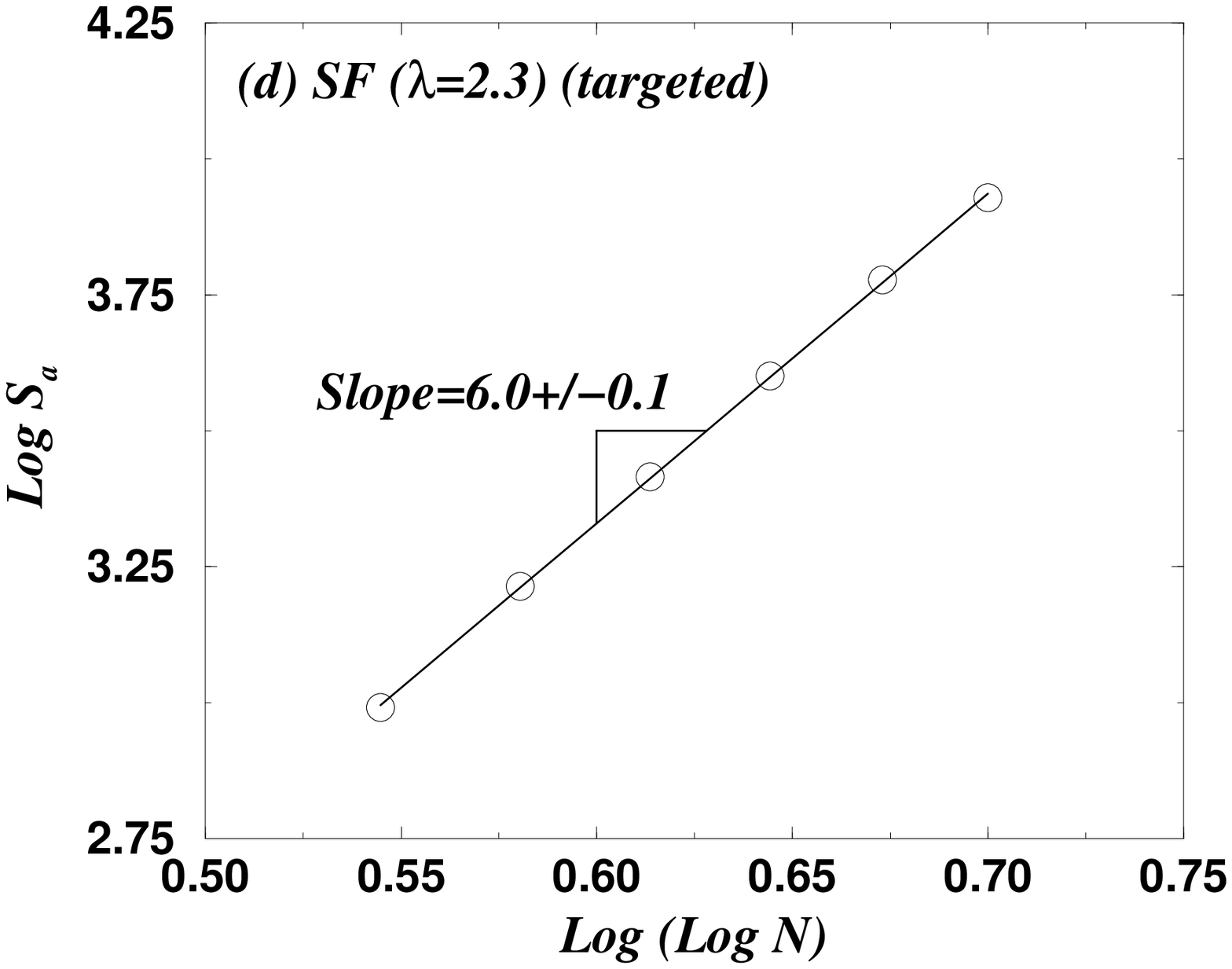,height=7cm,width=8cm}
\end{center}
\caption{Simulation results ((a) through (c)) for 
the scaling of $S_{a}/(c(p)N^{\delta})$ vs. 
$P_{\infty}N/(c(p)N^{\delta})$ for various types of networks, 
for random and targeted removal 
for $N$ between 1600 and 25600, and $a$ between 1.0 and 4.
(a) Erd\H{o}s-R\'{e}nyi networks (random) with $\langle k\rangle=3$ and 
$p=0.5,0.6$ and 0.7.
(b) Scale-free networks (random) with $\lambda=3.5$, $m=2$ 
and $p=0.6,0.7$ and 0.8. 
(c) Scale-free networks (targeted) for $\lambda=3.5$, $m=3$ and 
$a$ between 1.01 and 3.0, and targeted removal with 
$p=0.92$ and 0.94.
(d) Simulation results of $\log S_a$ vs. $\log\log N$ and comparison to the theoretical 
prediction (line) for $\delta$ for SF networks (targeted) for $\lambda=2.3$, $m=3$, $a=1.5$ 
and $p=0.97$.
}
\label{S_p_scaling_fig}
\end{figure}

Our results for the different regimes of $S_{a}$
can be summarized by the scaling relation for $p>p_c$
\begin{equation}
S_{a}\sim c(p)N^{\delta} f\left(\frac{P_{\infty}N}{c(p)N^{\delta}}\right)
\qquad \text{(Erd\H{o}s-R\'{e}nyi)},
\label{S_p_scaling}
\end{equation}
where  
$P_{\infty}$ is the probability of an arbitrary node to belong to 
the usual percolation spanning cluster~\cite{Percolation}.
The function $f(x)$ scales as $x$ when $x\ll 1$ and approaches a constant as $x\gg 1$. 
In Fig.~\ref{S_p_scaling_fig}(a), we present simulation results for several
$a$ and $p$ values for ER networks, supporting the scaling form of Eq.~(\ref{S_p_scaling}). 

The theory for LPP can be extended to all random networks 
with typical distance between nodes of order $\log N$ by substituting 
$\langle k\rangle$ with the generalized form $(\kappa -1)$,
known as the branching factor, defined by 
$\kappa-1 \equiv \frac{\langle k^2 \rangle}{\langle k \rangle}-1$~\cite{Cohen}. 
Replacing $\langle k \rangle$ with $(\kappa -1)$ in Eq.~(\ref{S_N_ER_delta}) we 
obtain the general equation for the spanning cluster size
\begin{equation}
S_a\sim c(p)(\kappa-1)^{a\frac{\log{N}}{\log{(\kappa_o-1)}}}=
N^{\delta}, \qquad
\delta\equiv a\frac{\log(\kappa-1)}{\log(\kappa_o-1)}
\label{S_N_delta}
\end{equation} 
where $\kappa_o-1$ is the branching factor of the original network
and $\kappa-1$ the branching factor after removal, which depends on $p$.
When a random fraction of the network is removed,
$\kappa-1=p(\kappa_o-1)$~\cite{Cohen}.
Note that for the specific case of ER networks, $\kappa-1=p\langle k \rangle$
and $\kappa_o-1=\langle k \rangle$, reducing Eq.~(\ref{S_N_delta}) 
to Eq.~(\ref{S_N_ER_delta}).
In the general case of random networks, 
the LPP transition is found by imposing $\delta=1$, which yields
\begin{equation}
\tilde{p}_c(a)=(\kappa_o-1)^{\frac{1-a}{a}}.
\label{tildep}
\end{equation}
The scaling form for $S_a$ is the same as Eq.~(\ref{S_p_scaling}) 
with $\delta$ taken from Eq.~(\ref{S_N_delta}).

Our general theory for LPP can be illustrated on scale-free (SF) networks. 
Scale-free networks have generated much interest due to their relation 
to many real-world networks, such as the Internet, 
WWW, social networks, cellular networks, and world-airline 
network~\cite{rev-Albert,BA,Amaral,Barrat,dyn-network}.
Scale-free networks are characterized by a power-law degree distribution 
$\Phi(k)\sim k^{-\lambda}$ ($m\le k\le K$), where 
$K\equiv mN^{1/(\lambda-1)}$~\cite{Cohen}.
The power-law distribution allows a network to have a few nodes
with a large number of links (``hubs'') which usually play a critical role
in network function. Calculating $\kappa$ for SF networks one obtains~\cite{Cohen}
\begin{equation}
\kappa = \left(\frac{2-\lambda}{3-\lambda}\right) \frac{K^{3-\lambda} 
- m^{3-\lambda}}{K^{2-\lambda} - m^{2-\lambda}}.
\label{kappa}
\end{equation}
For $\lambda>3$, Eq.~(\ref{S_N_delta}) is valid and thus
LPP is similar to ER networks, except that it depends on $\kappa-1$ instead of 
$\langle k \rangle$. 
The phase diagram of SF networks is shown in Fig.~\ref{phase_diagram}(b).
The results of the simulations supporting the theoretical value of 
$\delta$, Eq.~(\ref{S_N_delta}), are shown
in Fig.~\ref{S_a_p_N}(b), and for the scaling form of $S_a$ are presented 
in Fig.~\ref{S_p_scaling_fig}(b).

For $2 < \lambda < 3$ the typical network length scales 
as $\ell=2\log\log N/|\log(\lambda-2)|$~\cite{Cohen-2,math}.
For this regime, our scaling approach to calculate $S_a$ is no longer valid 
since the tree approximation breaks down.
However, the LPP transition still exists when $a\ell_{ij}=\ell_{ij}^{'}$,
where $\ell_{ij}^{'}$ is the distance after removal, with typical
value $\ell^{'}=2\log\log P_{\infty}N/|\log(\lambda-2)|$~\cite{fn_a_l_lprime}. 
Solving $a\ell_{ij}=\ell_{ij}^{'}$ for $N\rightarrow\infty$, we obtain
\begin{equation}
a = \frac{\ell'_{ij}}{\ell_{ij}} = \frac{\log\log P_{\infty}N}{\log\log N} 
\rightarrow 1 \qquad \text{(Scale-free $2<\lambda<3$)}.
\end{equation}
This implies that $\tilde{p}_c\rightarrow 0$ and thus, for any finite $p$,
$S_a$ is always of order $N$.
The results of the simulations presented in Fig.~\ref{S_a_p_N}(c) 
support our prediction.

Up to this point, we have only considered random removal of links.
Another kind of removal is targeted removal where the nodes with the 
largest degree are removed first~\cite{Cohen}.
This kind of removal is common in many real world scenarios such as denial of service
attacks on WWW and delays in airline hubs.

In scale-free networks, targeted removal of a fraction of $q$ nodes
with the largest degree can be treated as random removal of
$q'=q^{(2-\lambda)/(1-\lambda)}$ of the network links~\cite{Cohen}. After removal,
the maximum degree is given by $K'=mq^{1/(1-\lambda)}$.
For $\lambda >3$, making the substitutions $q\rightarrow q'$ and 
$K\rightarrow K'$ in Eq.~(\ref{S_N_delta}) we obtain the
equation for $\tilde{p}_c$~\cite{fn_pc_attack} and the scaling form for $S_a$ 
(see Table~\ref{table1}). 
The change to $q'$ and $K'$ reflects 
the fast collapse of the network and the rapid change in the typical network 
length. The transition line $\tilde{p}_c(a)$ in targeted removal decreases
significantly more slowly compared to random removal as seen 
in Fig.~\ref{phase_diagram}(b).   

In targeted removal for $2<\lambda<3$, removing even a small fraction
of the hubs produces a change in the distance from 
$2\log\log N/|\log(\lambda-2)|$ to $\log P_{\infty}N/\log(\kappa-1)$~\cite{Cohen-2,math}. 
Thus, after percolation $S_a$ can be calculated using the tree approximation
which yields
\begin{equation}
S_a\sim (\kappa-1)^{2a\frac{\log\log N}{|\log(\lambda-2)|}}=
(\log N)^{2a\frac{\log(\kappa-1)}{|\log(\lambda-2)|}}
\qquad\text{(Scale-free $2<\lambda<3$, targeted removal)}.
\label{S_lnN}
\end{equation}
In this case, the phase transition to a spanning cluster of order $N$ cannot be 
achieved for any finite value of $a$ and $p<1$, as seen from Eq.~(\ref{S_lnN}). 
Simulation results supporting Eq.~(\ref{S_lnN}) are shown in 
Fig.~\ref{S_p_scaling_fig}(d).
Comparing random to targeted removal for $2 < \lambda < 3$ for LPP yield
entirely opposite results. In random removal,
order $N$ nodes are still connected through the original paths. 
On the other hand, in targeted removal 
for any finite $a$, the network collapses into logarithmically small clusters. 

In summary, our results suggest that the usual percolation theory cannot 
correctly describe 
connectivity when only a limited set of path lengths are useful. In usual percolation, 
order $N$ of the network nodes are connected when $p > p_c$. However, in LPP, when 
$p_c<p<\tilde{p}_c$, only a zero fraction of the network is connected. Therefore, a 
much smaller failure of the network can lead to an effective network breakdown. 
As an illustration, consider an ER network with $\langle k\rangle=3$, 
and limit the length between nodes to $a=1.5$ times the original 
length. The theory of LPP predicts that the removal 
of $q=0.31$ of the network links is enough 
to break down the network, compared to $q=0.67$ in regular percolation. 
In the context of infectious diseases, if the virus typically survives 
up to $1.5\log N$ steps,
our theory predicts that the immunization threshold is significantly 
smaller, 0.31 compared to 0.67. 
Due to the above considerations, we expect our results to be important 
for network design, routing protocols and immunization strategies.
\bigskip\bigskip

\noindent
We thank the Israel Science Foundation, Israel Internet Association,
Yeshaya Horowitz Association and The Center for Complexity Science (Israel), 
the European NEST project DYSONET, DOE and ONR for financial support,
and J. C. Miller, E. Perlsman, and S. Sreenivasan for fruitful discussions.

\begin{table}
\begin{tabular}{|c||c|c|c|c|c|}\hline
Quantity&ER&\multicolumn{2}{c|}{SF random}&\multicolumn{2}{c|}{SF targeted}\\
\cline{3-4}\cline{5-6}
& &$2<\lambda<3$&$\lambda>3$&$2<\lambda<3$&$\lambda>3$\\\hline\hline
$\tilde{p}_c$&$\langle k\rangle^{(1-a)/a}$&0&$(\kappa_o-1)^{(1-a)/a}$
&1&$\tilde{p}_c(a,\kappa,\kappa_o)$\cite{fn_pc_attack}\\\hline
$S_a$&$N^{\delta}$&$N\hspace{0.2cm}(p>\tilde{p}_c)$&$N^{\delta}$
&$(\log N)^{\delta}$&$N^{\delta}$\\\hline
$\delta$&$a\left(1-\frac{|\log p|}{\log\langle k\rangle}\right)$
&1&$a\left(1-\frac{|\log p|}{\log(\kappa_o-1)}\right)$&
$2a\frac{\log(\kappa-1)}{|\log(\lambda-2)|}$&$a\frac{\log(\kappa-1)}{\log(\kappa_o-1)}$\\\hline
\end{tabular}
\caption{The functions $\tilde{p}_c$, $S_a$ and $\delta$ for several kinds
of network structures under random and targeted removal. The scaling of 
$S_a$ is given for $p<\tilde{p}_c$ except for scale-free networks
with $2<\lambda<3$ under random removal where $p$ is always above $\tilde{p}_c$.
}
\label{table1}
\end{table}

\bigskip                                                      
                                                              
\begin{thebibliography}{99}

\bibitem{rev-Albert} 
R. Albert and A.-L. Barab\'{a}si. Rev. Mod. Phys. \textbf{74}, 47
(2002); R. Pastor-Satorras and A. Vespignani, {\it
Structure and Evolution of the Internet: A Statistical Physics Approach}
(Cambridge University Press, Cambridge, 2004); S. N. Dorogovtsev and
J. F. F. Mendes, \emph{Evolution of Networks: From Biological Nets to
the Internet and WWW} (Oxford University Press, Oxford, 2003).

\bibitem{Wasserman} 
S. Wasserman, K. Faust, D. Iacobucci, M. Granovetter, {\it Social Network
Analysis: Methods and Applications} (Cambridge, 1994).

\bibitem{Pastor-Satorras} R. Pastor-Satorras and A. Vespignani
Phys. Rev. Lett. {\bf 86}, 3200-3203 (2001).

\bibitem{Newman} M. E. J. Newman, Phys. Rev. E {\bf 66}, 016128 (2002).

\bibitem{Lopez} E. L\'{o}pez et al.
Phys. Rev. Lett. {\bf 94}, 248701 (2005).

\bibitem{Sreenivasan} S. Sreenivasan, R. Cohen, E. L\'{o}pez, Z. Toroczkai, H. E. Stanley,
Phys. Rev. E (2007) (in press); cs.NI/0604023.

\bibitem{Barabasi} R. Albert, H. Jeong and A.-L. Barab\'{a}si, Nature {\bf 406}, 378 (2000).

\bibitem{Cohen} R. Cohen, K. Erez, D. ben-Avraham, and S. Havlin,
Phys. Rev. Lett. \textbf{85}, 4626 (2000).
{\it ibid.} {\bf 86}, 3682 (2001).

\bibitem{Callaway} D. S. Callaway, M. E. J. Newman, S. H. Strogatz, and D. J. Watts,
Phys. Rev. Lett. {\bf 85}, 5468 (2000).

\bibitem{Tanizawa} T. Tanizawa et al.
Phys. Rev. E {\bf 71}, 047101 (2005).

\bibitem{Percolation} S. Kirkpatrick, Rev. Mod. Phys. {\bf 45}, 574 (1973);
D. Stauffer and A. Aharony, {\it Introduction to Percolation Theory, 2nd ed.} 
(Taylor and Francis, 2004); \emph{Fractals and Disordered Systems}, 2nd. ed., edited by
A. Bunde and S. Havlin (Springer, Berlin 1996).

\bibitem{ER} P. Erd\H{o}s and A. R\'{e}nyi, Publ. Math. (Debrecen),
\textbf{6}, 290 (1959);  P. Erd\H{o}s and A. R\'{e}nyi, A. Publ. Math. 
Inst. Hung. Acad. Sci. {\bf 5}, 1760 (1960).

\bibitem{Bollobas} B. Bollob\'{a}s, \emph{Random Graphs} 
(Academic Press, Orlando, 1985).

\bibitem{Braunstein} L. A. Braunstein et al.
Phys. Rev. Lett. {\bf 91}, 168701 (2003).

\bibitem{fn_cp} For a tree of degree $z+1$ and depth $\ell$, the size is given by
$S_a=1+(z+1)\Sigma_{n=0}^{\ell-1} z^n\sim (\frac{z}{z-1})z^\ell.$
If $z=p\langle k\rangle$ the tree size scales as 
$[p\langle k\rangle/(p\langle k\rangle-1)]\left[p\langle k\rangle\right]^\ell$.

\bibitem{fn_network_algorithm}
To construct an Erd\H{o}s-R\'{e}nyi network, we begin with $N$ nodes
and connect each pair with probability $\phi$.
To generate a scale-free network with $N$ nodes, we use the 
Molloy-Reed algorithm \cite{Molloy-Reed}, which allows for the
construction of random networks with arbitrary degree distribution.  
We generate $k_i$ copies of each node $i$, 
where the probability of having $k_i$ 
satisfies $P(k_i)\sim k_i^{-\lambda}$.
These copies of the nodes are
then randomly paired in order to construct the network, making sure that
two previously-linked nodes are not connected again, and also excluding
links of a node to itself.

\bibitem{Molloy-Reed} M. Molloy and B. Reed, Random Struct. Algorithms
\textbf{6}, 161 (1995).

\bibitem{fn_sim} The size of $S_a$ was measured by: 
i) generating $R$ realizations of network structure, ii) performing 
percolation on each realization, 
iii) choosing $M$ nodes at random from the largest cluster, iv) 
for each node measuring the number of nodes connected satisfying 
$\ell'_{ij}\leq a\ell_{ij}$, and v) averaging over the results.
Typically, $R=1000$ and $M=100$.

\bibitem{BA} A.-L. Barab\'{a}si and R. Albert, Science
\textbf{286}, 509 (1999). H. A. Simon, Biometrika {\bf 42}, 425 (1955).

\bibitem{Amaral} L. A. N. Amaral, A. Scala, M. Barth\'{e}l\'{e}my, and H. E. Stanley,
Proc. Nat. Acad. Sci. USA {\bf 97} 11149 (2000).

\bibitem{Barrat} A. Barrat et al. 
Proc. Nat. Acad. Sci. USA, {\bf 101}, 3747 (2004).

\bibitem{dyn-network} P. L. Krapivsky, S. Redner, and F. Leyvraz,
Phys. Rev. Lett. \textbf{85}, 4629 (2000).

\bibitem{Cohen-2} R. Cohen and S. Havlin Phys. Rev. Lett. {\bf 90}, 058701 (2003). 

\bibitem{math} R. van der Hofstad, G. Hooghiemstra, D. Znamenski, math.PR/0502581 (2005).

\bibitem{fn_a_l_lprime} Indeed, for $\lambda>3$, $a\ell_{ij}=\ell_{ij}^{'}$ 
reduces to Eq.~(\ref{S_N_delta}).

\bibitem{fn_pc_attack} The function $\tilde{p}_c(\kappa,\kappa_o)$ is given implicitly
by solving $\delta=1$ from Eq.~(\ref{S_N_delta}), using $\kappa$ from Eq.~(\ref{kappa})
with $q=q'$ and $K=K'$ for targeted removal.

\end{thebibliography}
\end{document}